# Scavenging effect on plasma oxidized $Gd_2O_3$ grown by high pressure sputtering on Si and InP substrates

M A Pampillón[1], P C Feijoo[1], E San Andrés[1], H García[2], H Castán[2] and S Dueñas[2]

1 Dpto. Física Aplicada III: Electricidad y Electrónica. Universidad Complutense de Madrid. Facultad de Ciencias Físicas. Avda. Complutense S/N, 28040, Madrid, Spain

2 Dpto. Electricidad y Electrónica, E.T.S.I. Telecomunicación, Universidad de Valladolid, 47011, Valladolid, Spain

**Abstract**

In this work, we analyze the scavenging effect of titanium gates on metal-insulator-semiconductor capacitors composed of gadolinium oxide as dielectric material deposited on Si and InP substrates. The $Gd_2O_3$ film was grown by high pressure sputtering from a metallic target followed by an in situ plasma oxidation. The thickness of the Ti film was varied between 2.5 and 17 nm and was capped with a Pt layer. For the devices grown on Si, a layer of 5 nm of Ti decreases the capacitance equivalent thickness from 2.3 to 1.9 nm without compromising the leakage current ($10^{-4}$ A cm$^{-2}$ at $V_{gate}$ =1 V). Thinner Ti has little impact on device performance, while 17 nm of Ti produces excessive scavenging. For InP capacitors, the scavenging effect is also observed with a decrease in the capacitance equivalent thickness from 2.5 to 1.9 nm (or an increase in the accumulation capacitance after the annealing from ~1.4 to ~1.7–1.8 µF cm$^{-2}$). The leakage current density remains under $10^{-2}$ A cm$^{-2}$ at $V_{gate}$ =1.5 V. For these devices, a severe flatband voltage shift with frequency is observed. This can be explained by a very high interface trap state density (in the order of $10^{13}$–$10^{14}$ eV$^{-1}$ cm$^{-2}$).

## 1. Introduction

Several high permittivity (κ) dielectrics are currently being studied in order to continue with Si-based metal-oxidesemiconductor field effect transistors (MOSFETs) downscaling [1–3]. Besides, other applications using high mobility or high bandgap semiconductors need these high κ materials in order to improve the performance of the devices [4–6]. The main challenge in MOSFETs applications is to prevent the low κ oxide growth at the dielectric/semiconductor interface with the aim of obtaining a low capacitance equivalent thickness (CET) (or a low equivalent oxide thickness, EOT, which is ∼0.7 nm smaller that the CET) without compromising the leakage current of the device and the interface trap state density, $D_{it}$. For Si substrates, the interfacial $SiO_x$ has reasonable electrical properties and is manufacturing friendly, but for thicknesses under ∼1 nm the mobility degrades [1]. Also, with high κ oxides it is difficult to control these thin interfacial oxides [1–3]. The thickness of the interfacial $SiO_x$ imposes a lower limit on CET that must meet the requirements for future MOSFETs devices [7]. Concerning the native oxides of other semiconductors, they do not have as good properties as $SiO_x$ (for instance, $GeO_x$ is water soluble and thermally unstable [8] and no oxidation process has been published to fabricate well behaved MOS structures on III–V semiconductors). For those reasons, obtaining a low $D_{it}$ with these alternative semiconductors is a great challenge that must be eventually solved [9].

This paper studies the scavenging concept [10], aiming to reduce the CET value on several kinds of semiconductors. Kim et al [10] explained the scavenging effect observing how capping layers of reactive metals have the property of acting as oxygen-gettering and reducing the $SiO_x$ interlayer formed between the semiconductor and the metal oxide film. In fact, they focused on Ti, which is a well known reactive metal that is complementary MOS (CMOS) compatible. Its surface tends to oxidize due to its negative redox potential of −1.63 V [11]. This Ti property is used, for instance, to improve vacuum in the Ti sublimation pumps, where the pressure is reduced by deposition of thin Ti layers in the chamber. The clean Ti surface absorbs the residual gas, mostly composed of water molecules, improving residual vacuum. Kim et al demonstrated that Ti works as interfacial $SiO_x$ scavenger both with $HfO_2$ and $ZrO_2$.

In this work, we used gadolinium oxide as the high κ material and as the gate electrode, several thicknesses of Ti capped with Pt. The devices were fabricated both on Si and InP substrates. The scavenging effect has already been reported on Si [12] and also on Ge [13], but, as far as we know, no one has published results on III–V semiconductors. Thus, it is very important to check the applicability of the interface scavenging process to this family of semiconductors, since most microelectronic companies are exploring high mobility III–V materials as the channel semiconductor for their future n-MOS devices.

$Gd_2O_3$ presents good properties [14] on Si [15], on SiGe [16], on alternative substrates as III–V semiconductors [17] and also on high bandgap materials [18]. In this work, $Gd_2O_3$ was deposited on Si and InP substrates by means of the high pressure sputtering (HPS) technique. The high working pressure of this non-conventional system (three orders of magnitude above conventional sputtering systems) prevents damaging the semiconductor surface due to the low thermalizing length of the particles: at a pressure of 0.5 mbar, the mean free path of the sputtered particles, between 1–3 mm, is much smaller than the target-substrate distance, ∼2.5 cm. Our group has previously used this system to deposit several high κ dielectrics from oxide targets: $TiO_2$ [19], $HfO_2$ [20], $Sc_2O_3$ [21], $Gd_2O_3$ [22] and also, ternary oxides as $GdScO_3$ [23]. The novelty of the $Gd_2O_3$ deposition used in this work follows a work reported by Hoshino et al [24] and consists in the growth of the dielectric film from a metallic Gd target by a two-step process: firstly we deposited a thin film of Gd and then performed, without breaking the vacuum, an $Ar/O_2$ plasma exposure in order to oxidize the Gd layer. This process avoids semiconductor surface oxidation and degradation since it is not exposed to an oxygen ambient when the sample is into the chamber. The details are published in [25]. After insulator deposition, a Ti/Pt stack was e-beam evaporated to form the top contact of the MOS capacitor. Several thicknesses of Ti were used (from 2.5 to 17 nm) in order to study the Ti thickness influence on scavenging. We also studied the effects of thermal annealing on the electrical characteristics of the devices by using several annealing conditions.

## 2. Experimental

The substrates used in this work were 2″ n-type Si wafers and 2″ wafers of undoped InP (so, n-type behavior). Both kinds of wafers had (100) orientation and were single side polished with a resistivity of 1.5–10 Ω cm for Si and 1–5 Ω cm for InP. Before processing, a 200 nm $SiO_x$ field oxide was evaporated to isolate the different capacitors. The device sizes ranged from 50×50 to 700×700 $\mu m^2$. Standard RCA (Radio Corporation of America) cleaning [26] was performed for Si samples, while InP substrates were cleaned with 10% diluted iodic acid for 1 min [27]. Before introducing the samples into the chamber, a 30 s HF dip was carried out in order to remove the native oxides. In this work, we made three sets of samples, growing $Gd_2O_3$ films by HPS with a two step-process [25, 28]. For the first set, a film of Gd was deposited in an Ar atmosphere for 80 s on Si. Afterwards, without breaking the vacuum, a long plasma oxidation is performed by introducing a 5% of oxygen flow into the chamber for 300 s. Both steps were carried out at room temperature with a pressure of 0.5 mbar and 30 W of radiofrequency (rf) power. The second set of samples also have a metallic Gd deposition of 80 s on Si and a milder oxidation step, 100 s of plasma oxidation at only 20 W. Finally, a third set of samples was fabricated on InP. Since the CET requirement is less strict on III–V semiconductors due to the higher mobility, we kept the same oxidation conditions but deposited a thicker metallic Gd layer by extending the first step to 120 s. This should produce a 50% thicker $Gd_2O_3$ film.

For the top contact, in the first set of samples, thick Pt and thick Ti electrodes were e-beam evaporated and defined by lift-off. Forming gas annealing (FGA) at 300 °C for 20 min was performed. In the second set, after $Gd_2O_3$ deposition, the Si wafers were divided in four pieces to ensure that all samples have identical dielectric film. We studied Ti thicknesses of 2.5, 5 and 17 nm. All the Ti layers were in situ capped with Pt in order to avoid the nitridation and/or oxidation of the metal gate. Finally, given the Si results, for InP, we focused on a Ti thickness of 5 nm. For the Si samples, standard FGAs were performed during 20 min at 300, 350 and 400 °C consecutively. To minimize phosphorous loss, the FGA temperature of the InP devices was limited to 325 °C. Thus, to ensure a good ohmic contact of the back electrode, the FGA was extended to 30 min.

High resolution transmission electron microscopy (HRTEM) images of the MIS capacitors of the first set of samples were obtained using a Tecnai T20 microscope from FEI, operating at 200 keV. Capacitance and conductance versus gate voltage (C–$V_{gate}$ and G–$V_{gate}$) curves were measured at frequencies from 100 Hz to 10 MHz with an Agilent 4294A before and after the FGAs. Also, the leakage current density as a function of gate voltage (J–$V_{gate}$) was obtained with a Keithley 2636A and a Keithley 6517A programmable electrometer. To determine interface quality (or density of defects), deep level transient spectroscopy (DLTS) measurements were carried out with a Boonton 72B capacitance meter, an HP54501 digital oscilloscope to record the capacitance transients and an HP81104 pulse generator to apply cool down the temperature from room temperature to 77 K.

3. **Results and discussions**

The cross-sectional HRTEM images of the sample with 80 s of Gd and a 300 s of plasma oxidation at 30 W after FGA at 300 °C are shown in figure 1. Figure 1(a) shows the sample with Pt as top metal (as a reference) and figure 1(b), the same dielectric film but with a thick Ti layer. The pure Pt sample has a 6 nm of an amorphous $Gd_2O_3$ layer and 1.5 nm of interfacial $SiO_x$, probably grown during the long plasma oxidation. For that reason, less aggressive plasma oxidation (with less duration and lower rf power) was carried out in the second set of samples.

On the other hand, figure 1(b) shows that when Ti is used as top metal, no $SiO_x$ interface can be observed. This is due to the scavenging effect of Ti. It is also noticeable, that the thickness of the $Gd_2O_3$ film is 2 nm lower than in the Pt case. This result points to an excessive scavenging of the thick Ti layer that not only removes oxygen from the $SiO_x$ interface, but also scavenges part of the dielectric material

The effects of this excessive scavenging can be observed in figure 2, where the C–$V_{gate}$ curves for these samples are presented. For the Pt sample (figure 2(a)), the accumulation capacitance before and after the FGA at 300 °C does not change significantly. However, the Ti sample (figure 2(b)) exhibits an increase in the accumulation capacitance value after the FGA at 300 °C, which can be explained by the scavenging effect we have observed in figure 1(b). The distortion in the accumulation capacitance for Ti is related to a huge increase of gate leakage (not shown here). In fact, higher annealing temperatures produce even higher leakage, yielding non-functional devices. This means that the scavenging effect has to be controlled, either by limiting the thermal budget, or by reducing the thickness of the Ti layer. After these results and following [12], smaller thicknesses of Ti were also studied with the aim to control the scavenging effect.

To optimize the scavenging effect, aiming for the decrease of the interfacial $SiO_x$ thickness without degrading $Gd_2O_3$, a second set of devices was fabricated (80 s of Gd and 100 s of a plasma oxidation at 20 W). Since the metallic Gd deposition time is the same as in the first set, we can assume that the $Gd_2O_3$ thickness is the same, 6 nm. On the other hand, we expected a thinner interface since we used a less aggressive oxidation. Figure 3 shows the normalized C–$V_{gate}$ characteristic of the second set of samples with different thicknesses of Ti, measured at 10 kHz. For 2.5 nm of Ti (figure 3(a)), and before annealing the sample, the C–$V_{gate}$ curve shows a characteristic hump in depletion due to the $D_{it}$. This distortion is observed for every Ti thickness before FGA. After annealings, there is a slight increase of the accumulation capacitance. Also, there are no relevant differences between the several temperatures of the FGAs. Then, the scavenging effect of the 2.5 nm Ti film is moderate and saturates for an annealing temperature of 300 °C. Furthermore, the FGAs make the C–$V_{gate}$ fall more abruptly and the hump in depletion disappears, which indicates an improvement in the oxide/ semiconductor interface. For the 5 nm Ti (figure 3(b)), the accumulation capacitance value visibly increases after the FGA. No significant differences can be noticed between the FGA at 350 and 400 °C for this sample. In this case, the scavenging saturation occurs for the FGA at 350 °C. For this thickness, the distortion of the curve in depletion is reduced after annealing, but it is still clearly observed. For the thicker Ti layer (figure 3(c)), the behavior of the C–$V_{gate}$ curves is similar to the former set up to an annealing temperature of 350 °C, but for annealing temperatures above 400 °C, there is a severe accumulation capacitance drop, pointing to an excessive scavenging effect. The normalized conductance for this sample presents a value over 1 S cm$^{-2}$ for gate voltages higher than 1 V (not shown here), supporting this aggressive scavenging effect.

Figure 4 presents the tendency of the CET value as a function of the annealing temperature for the second set of samples. For the thinner Ti thickness, there is a hardly noticeable decrease in the CET from a value slightly higher than 2.3 nm (for the as-deposited sample) to 2.2 nm (after FGAs). When comparing this value with the Pt gated CET (around 2.6 nm [25]), we can conclude that some scavenging takes place even during the deposition of the metal contact. The saturation effect for this sample commented on in the former paragraph is also noticed in this figure for annealing temperatures above 300 °C. For the sample with 5 nm Ti layer, the CET reduction goes from a value slightly lower than 2.3 nm (unannealed sample) to 1.9 nm (after FGA at 400 °C). Finally, for the sample with the thicker Ti top metal, the trend in the CET diminution is the same as observed in the previous sample until FGA at 350 °C. Due to the capacitance drop observed in figure 3(c)) after FGA at 400 °C, it is not possible to obtain the CET value for this temperature. Besides, from figure 4, we observe that the CET is similar for all the as-deposited samples, but is lower for samples with higher Ti thickness. This is also an indication that during the top contact evaporation there is some scavenging in the samples, most likely due to heating by infrared radiation.

Figure 5 represents the trend in the $D_{it}$ (obtained with the conductance method [29]) as a function of the annealing temperature for the different thicknesses of Ti. For the sample with 2.5 nm of Ti, the $D_{it}$ achieves a low value of $10^{11}$ eV$^{-1}$ cm$^{-2}$ after the FGA at 300 °C, showing an interface improvement, as was also qualitatively observed in the C–$V_{gate}$ of that sample. However, thicker Ti layers show a higher value of interfacial trap density even before annealing, which is another confirmation of the scavenging effect during evaporation (lower unannealed CET means more intensive scavenging, which produces more defects). For these samples, the lowest FGA temperature, 300 °C, produces a reduction of the $D_{it}$ up to $4$–$5 \cdot 10^{11}$ eV$^{-1}$ cm$^{-2}$. These values increase to the $10^{12}$ eV$^{-1}$ cm$^{-2}$ range when the annealing temperature is raised. $D_{it}$ degradation is even more noticeable for the sample with 17 nm of Ti after the FGA at 400 °C, suggesting again that excessive scavenging results in a defective interface. Other works have reported similar values of the $D_{it}$ using ZrO$_2$ [30], single crystalline Gd$_2$O$_3$ [31] and polycrystalline Gd$_2$O$_3$ with an amorphous GdSiO layer [32].

In figure 6, the leakage current density is represented as a function of the gate voltage. These results are in agreement with the capacitance measurements: the leakage of the sample with 2.5 nm of Ti does not change before and after FGAs, with a value around $10^{-7}$ A cm$^{-2}$ at 1 V (figure 6(a)). This means that only a negligible scavenging is happening. For the sample with 5 nm of Ti, current density increases moderately as the annealing temperature is raised. In any case, leakage current is in the order of $10^{-4}$ A cm$^{-2}$ at 1 V for all the FGAs (figure 6(b)), similar to those reported in previous works [30, 33]. On the other hand, for the sample with 17 nm of Ti, the current density reaches a high value over 0.1 A cm$^{-2}$ at 1 V after the FGA at 350 °C as can be observed in figure 6(c). This confirms that there is excessive scavenging in this sample.

Summarizing these results, 5 nm of Ti together with FGA at 300 °C is the best compromise between scavenging, $D_{it}$ and leakage current. However, more intense scavenging should not be completely discarded, but it would require a metal gate-last process. In other words, after FGA the Ti gate should be substituted by a threshold voltage ($V_{th}$) control metal followed by an interface improvement process (for instance, FGA at 500 °C during 20 min).

The frequency dispersion of the C–V$_{gate}$ curves of the sample with 5 nm of Ti and annealing at 400 °C is shown in figure 7. All measured frequencies (from 1 kHz to 1 MHz) present almost the same value of the accumulation capacitance, except the one at 1 MHz, which is around 10% lower than the others. This reduction in the capacitance is due to combined effect of the series resistance with a high conductance (over ~1 S cm$^{-2}$ at gate voltages above 0 V) measured at this high frequency. The hump of the capacitance in depletion due to the interface traps decreases when increasing frequency. The $D_{it}$ values obtained from this figure by using the conductance method decrease almost one order of magnitude as the frequency is increased (from $2\cdot 10^{12}$ to $4\cdot 10^{11}$ eV$^{-1}$ cm$^{-2}$). This means that the traps can follow the ac signal only at a moderately high frequency. We observed previously the same behavior in similar samples but using pure Pt as top metal [25]. That can be related to the existence of a border trap distribution inside the dielectric. As border traps are located further away from the interface, emission and capture time constants exponentially decrease with the distance from the interface [34]. Therefore, only traps at the interface contribute to conductance values at high frequency. Besides, DLTS gives a $D_{it}$ value around $10^{12}$ eV$^{-1}$ cm$^{-2}$ uniform through the gap (not shown here). These results are in good agreement with the values provided by the conductance method.

The C–V$_{gate}$ hysteresis characteristics at 10 kHz are presented in figure 8 before and after the FGA at 400 °C. The curve is obtained starting in accumulation. The flatband voltage (V$_{FB}$) shift is around 500 mV for the as-deposited sample and it is reduced to less than 50 mV after the FGA. These results are similar to other works reported [31, 35] and indicate that the FGA passivates most of the defects inside the Gd$_2$O$_3$ film that act as slow traps.

To complete the electrical study, the J–V$_{gate}$ at different temperatures is obtained to characterize the leakage mechanism. In figure 9(a), two regions can be distinguished. At low voltages (V$_{gate}$ <0.1 V), current density does not depend on the temperature. Therefore, tunneling is the dominant conduction mechanism in this region. In contrast, for higher voltages, current is thermally activated. This dependency fits well to the Poole–Frenkel effect, that is, trap assisted conduction mechanism is dominant at electric field values higher than 0.7 MV cm$^{-1}$. Figure 9(b) shows the plot of J/E (in logarithmic scale) against E$^{1/2}$ at several temperatures, corresponding to the sample with 5 nm of Ti layer and FGA at 400 °C. There is a linear dependence in the high-field range, as required by the Poole–Frenkel equation:

$$I = I_0 \exp\left(\frac{\beta_{FE} E^{\frac{1}{2}}}{kT}\right) E \qquad (1)$$

where I is the current, $I_0$, a pre-exponential factor, $\beta_{PF}$ is the Poole–Frenkel coefficient, E, the applied electric field, k is the Boltzmann's constant and T, the temperature.

The obtained value of $\beta_{PF}$ in the 0.7–1 MV cm$^{-1}$ electric field range slightly varies with temperature in the range $(0.7–1.3)\times 10^{-5}$ e V m$^{1/2}$ V$^{-1/2}$. Similar values were obtained in different MOS samples with $Gd_2O_3$ fabricated with the same method [36].

Finally, after the characterization of the Si, we applied the scavenging effect using InP as the semiconductor substrate and 5 nm of Ti capped with Pt as metal gate stack. CET downscaling of these devices is not so critical for high electron mobility substrates, so, for this study, we have grown a thicker $Gd_2O_3$ layer to ensure lower leakage current. According to the C–V$_{gate}$ curves shown in figure 10, accumulation capacitance increases clearly after the FGA (from ∼1.4 to ∼1.7–1.8 µF cm$^{-2}$ at 1.5 V for both measuring frequencies). This means a reduction in CET from 2.5 to 1.9 nm. In our previous works using Pt as top metal [37, 38], this effect was not observed after the FGAs, so the capacitance increase cannot be related to a change in the thickness of the dielectric material or to a permittivity increase. We can observe that the accumulation capacitance value does not change with frequency. In [39] it was shown that the combined effect of series resistance and conductance can produce the incorrect determination of capacitance. Also, a high $D_{it}$ can produce a capacitive signal that could be mistakenly interpreted as gate capacitance [40]. However, these problems can be detected by varying the measuring frequency: if the accumulation capacitance value changes, then the C–V$_{gate}$ curve has to be taken with care. Since this is not the case, we can be quite sure that the measured capacitances are due to the gate dielectric and correctly measured. Thus, we can associate the accumulation capacitance improvement to a reduction of the total thickness of the dielectric due to the scavenging effect of Ti in III–V, as was also found on Si. Concerning gate leakage (not shown here), there is a slight increase (less than 1 order of magnitude) in the current density after the FGA (from $10^{-3}$ A cm$^{-2}$ to less than $10^{-2}$

A cm$^{-2}$ at 1.5 V). This is also consistent with the reduction of the total dielectric thickness due to the scavenging effect.

From the measured normalized conductance (not shown here) and using the conductance method, the D$_{it}$ value obtained is very high, in the order of $10^{13}$ eV$^{-1}$ cm$^{-2}$, but comparable with the one obtained with pure Pt gates [37] and with other works of Al$_2$O$_3$/InP capacitors [41]. It should be noticed that, in this study, no interfacial treatments were applied to the InP substrate before the HPS process, such as nitrogen plasma exposure [42] or Si capping [43].

Also, from figure 10, a severe V$_{FB}$ shift is observed with the frequency (around 0.9 V for these two frequencies). To focus on this effect, the frequency dispersion of the C–V$_{gate}$ is presented in figure 11, measured from 100 Hz to 10 MHz. According to [44] and [45], the C–V$_{gate}$ frequency dependence can be attributed to a high amount of interface traps. Following these references, there is a logarithmic relationship between V$_{FB}$ and frequency, following the next equation:

$$V_{FB}(f) = -\frac{Q_f}{C_{ox}} \mp \frac{kTqD_{it}}{C_{ox}} \mathrm{Ln}\, f \quad (2)$$

where Q$_f$ is the fixed oxide charge, k the Boltzmann constant, q the elementary charge, T the temperature, and f the frequency measured in Hz. C$_{ox}$ is the normalized oxide capacitance and can be extracted from the accumulation value of the capacitance. The $\mp$ sign stays for acceptor or donor type interface traps, respectively. From this formula, the interface defect density D$_{it}$ is obtained.

Figure 12 represents the V$_{FB}$ data as a function of frequency. We observe that under 3 kHz, the fatband voltage remains constant, but above this frequency there is a remarkable linear dependence. Halova et al [45] explain this behavior by a characteristic response time of traps, that respond at low frequencies but only partially when the frequency increases. The linear fit for frequencies above 10 kHz is also shown. Using equation (2), the calculated donor trap density D$_{it}$ is $\sim 10^{14}$ eV$^{-1}$ cm$^{-2}$, one order of magnitude higher than that value obtained by the conductance method. In any case, the D$_{it}$ is too high for MOSFETs applications and a reduction of this value is required. This could be achieved with passivation treatments as Si interfacial passivation, surface sulfidation, in situ fluorination, etc [43, 46, 47]. In the near future we will focus on this interface improvement. Also, from the fit, the fixed charge obtained is $\sim 8 \cdot 10^{-6}$ C cm$^{-2}$.

## 4. Conclusions

The scavenging effect has been studied for $Gd_2O_3$ films deposited by HPS from a metallic Gd target followed by plasma oxidation. This effect has been studied on Si and InP substrates. A proper thickness of the Ti layer enables the reduction of the interfacial oxide without compromising the leakage of the devices. In Si substrates the optimum conditions were obtained for a Ti thickness of 5 nm and FGA at 300 °C. These devices presented a reduction of 0.4 nm of CET with a low leakage current of $\sim 10^{-4}$ A cm$^{-2}$ at 1 V, and a $D_{it}$ of $4 \cdot 10^{11}$ eV$^{-1}$ cm$^{-2}$, comparable with other high κ materials. On the other hand, the InP capacitors showed an increase in the accumulation capacitance that provides a CET lower than 2 nm. The current density for these InP devices is in the order of $10^{-2}$ A cm$^{-2}$, but the $D_{it}$ obtained is extremely high ($10^{13}$–$10^{14}$ eV$^{-1}$ cm$^{-2}$). Finally, we can conclude that although surface passivation treatment is needed to reduce $D_{it}$ in InP substrates, gate dielectric scavenging is possible with III–V semiconductors by using an optimized Ti/Pt electrode, as we have shown in this work with InP.

### Acknowledgments

The authors would like to acknowledge the collaboration of 'C.A.I de Técnicas Físicas' of 'Universidad Complutense de Madrid' and 'Instituto de Nanociencia de Aragón' of the 'Universidad de Zaragoza'. This work was financed by projects TEC2010-18051 and TEC2011-27292-C02-01 and FPI grant (BES-2011-043798) of the Spanish 'Ministerio de Economía y Competitividad'.

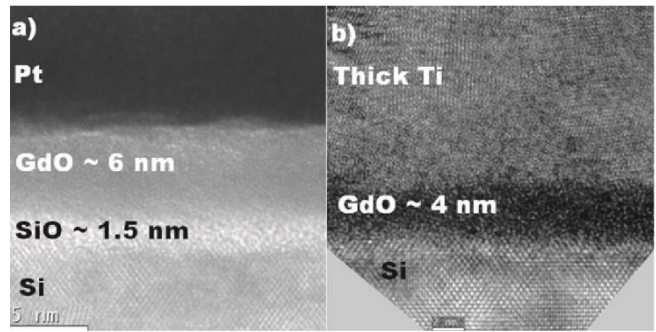

Figure 1. Cross-sectional HRTEM images of sample with 80 s of Gd and a 300 s plasma oxidation at 30 W after FGA at 300 C with (a) Pt and (b) thick Ti as top electrode.

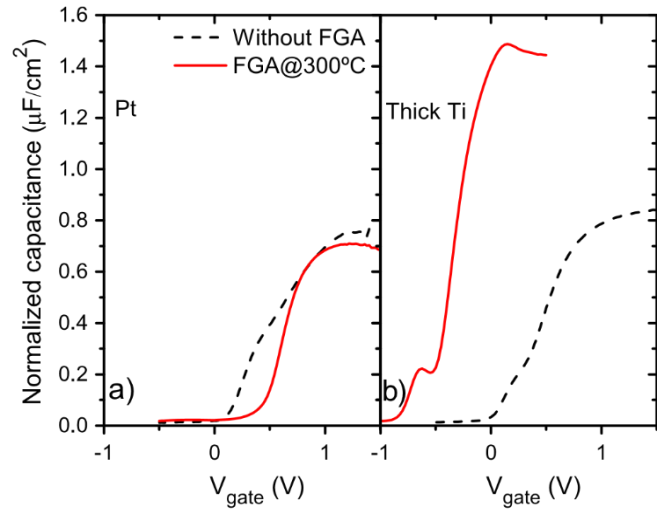

Figure 2. Area normalized C–V$_{gate}$ curves for the Si sample with 80 s of Gd and a 300 s plasma oxidation. The rf power was 30 W for both processes. The capacitance was measured at 10 kHz before and after the FGAs with (a) Pt and (b) thick Ti as top metal.

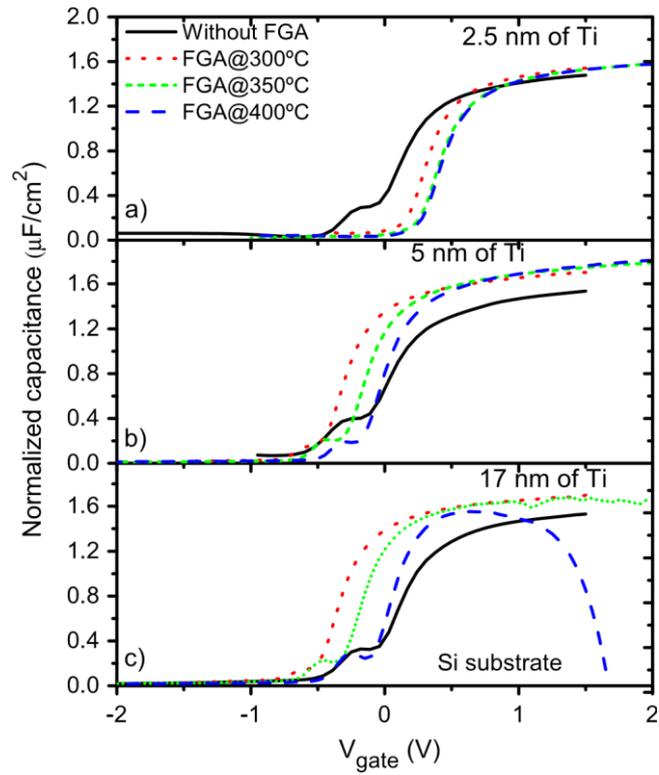

Figure 3. Area normalized C–V$_{gate}$ curves for the Si sample with 80 s of Gd and a 100 s plasma oxidation at 20 W measured at 10 kHz before and after the FGAs with different thicknesses of Ti: (a) 2.5 nm, (b) 5 nm and (c) 17 nm.

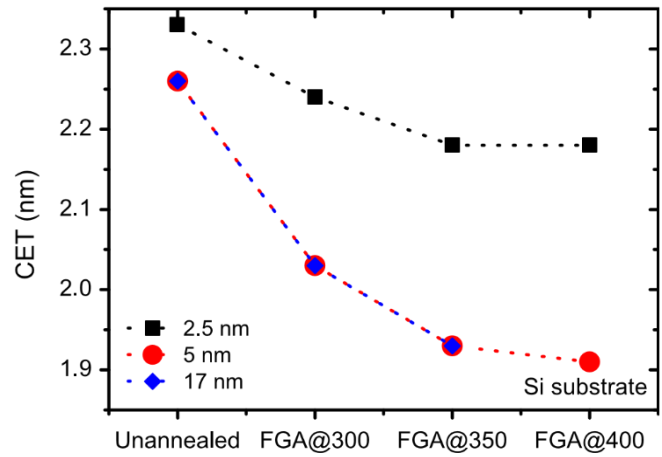

Figure 4. CET value as a function of the annealing temperature for Si sample with 80 s of Gd and a 100 s plasma oxidation at 20 W for different thicknesses of Ti.

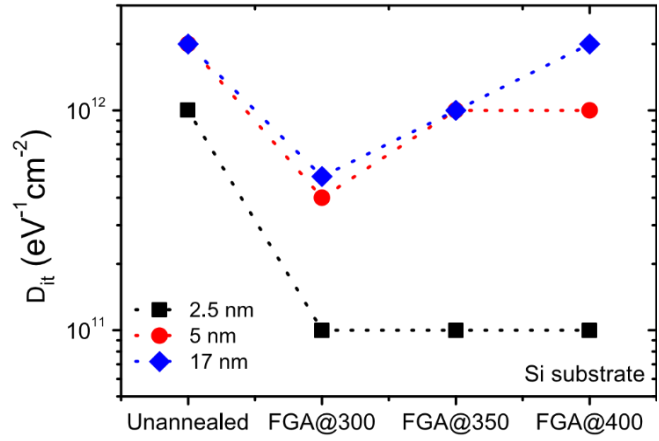

Figure 5. $D_{it}$ value obtained by the conductance method as a function of the annealing temperature for the different thicknesses of Ti.

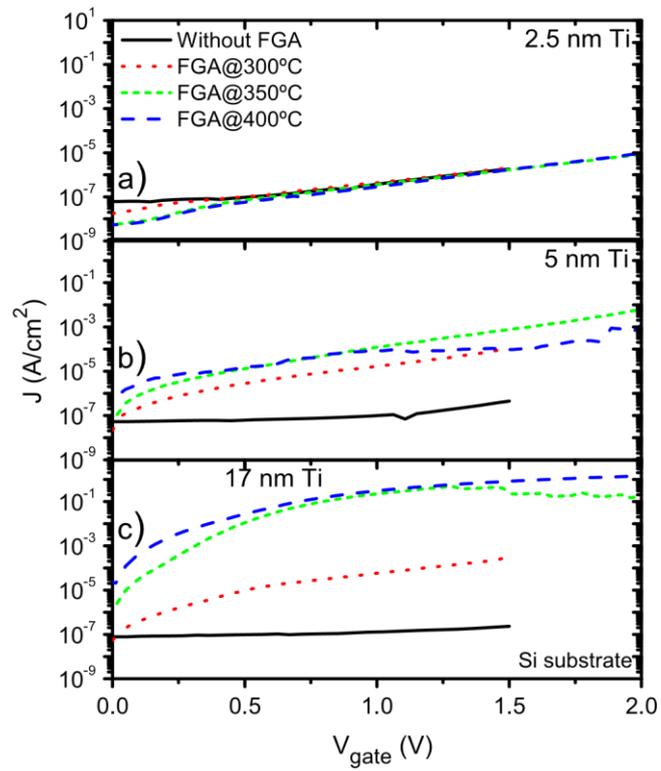

Figure 6. J–V$_{gate}$ characteristics for the Si sample with 80 s of Gd and a 100 s plasma oxidation at 20 W measured before and after the FGAs with different thicknesses of Ti: (a) 2.5 nm, (b) 5 nm and (c) 17 nm.

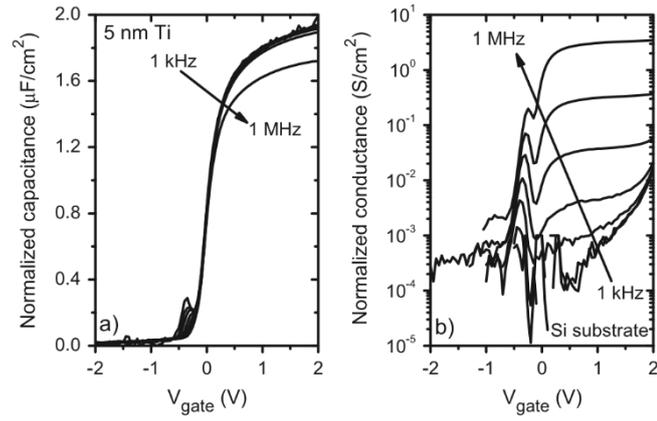

Figure 7. Frequency dispersion (a) C–$V_{gate}$ and (b) G–$V_{gate}$ curves of the Si sample with 80 s of Gd and 100 s of plasma oxidation with 5 nm of Ti after FGA at 400 °C.

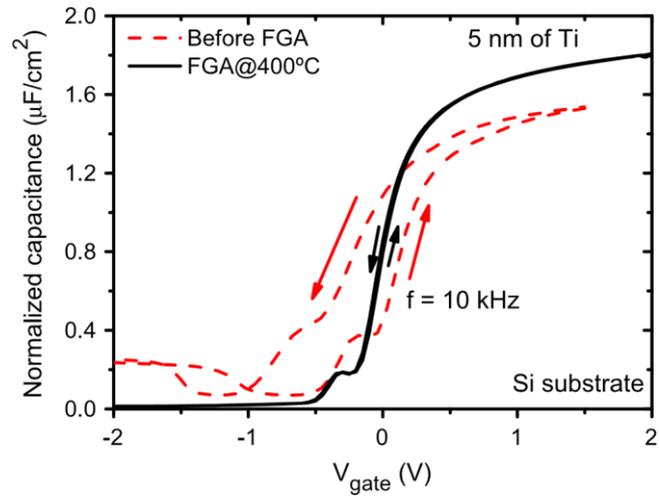

Figure 8. Hysteresis C–V$_{gate}$ characteristic measured at 10 kHz with a sweep from accumulation to inversion and back again for the Si sample with 5 nm of Ti before and after the FGA at 400 °C.

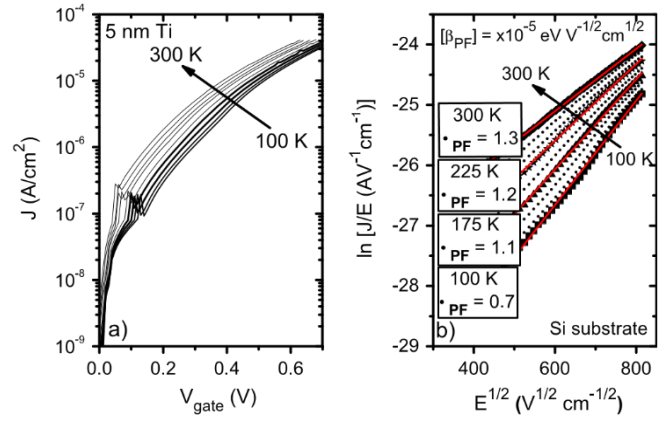

Figure 9. (a) J–V$_{gate}$ characteristics measured at different temperatures (from 100 K to 300 K) and (b) current electric field dependency fitting following the Poole–Frenkel model at several temperatures for the Si sample with 5 nm of Ti and after the FGA at 400 °C. The β$_{PF}$ parameter is shown.

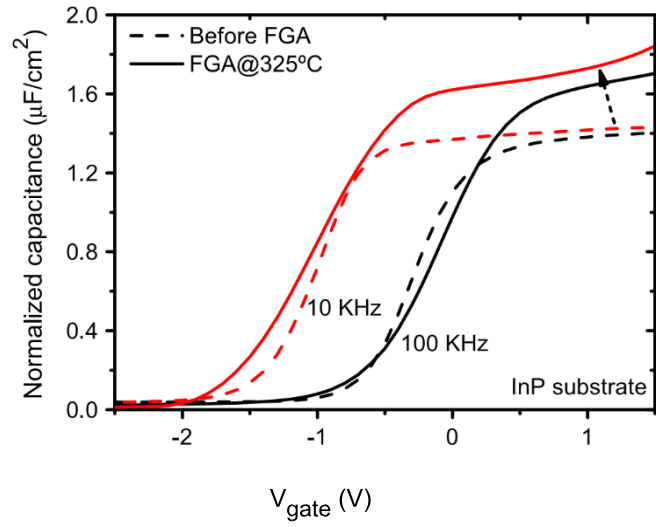

Figure 10. Normalized C–V$_{gate}$ for the InP sample with 120 s of Gd and 100 s of plasma oxidation at 20 W measured at 10 and 100 kHz with 5 nm of Ti before and after the FGA.

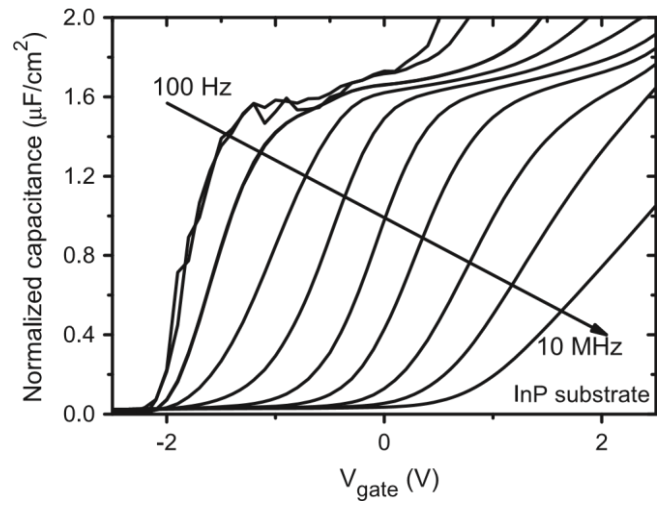

Figure 11. Frequency dispersion C–V$_{gate}$ curves for the InP sample after FGA at 325 °C for 30 min.

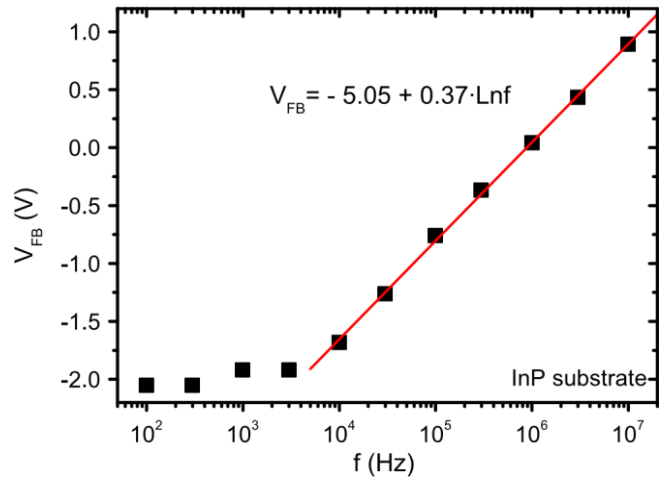

Figure 12. Experimental data for $V_{FB}$ as a function of the frequency measured for the InP sample after the FGA at 325 °C and linear fit.